\documentclass[a4paper]{article}
\usepackage{graphicx}
\usepackage{amsmath}
\usepackage{hyperref}
\usepackage{color}

\begin{document}

\begin{center}

{\bf \Large The norm game - how a norm fails}\\[5mm]

{\large Krzysztof Ku{\l}akowski$^+$ and Antoni Dydejczyk$^*$}\\[3mm]

{\em

Faculty of Physics and Applied Computer Science,

AGH University of Science and Technology,

al. Mickiewicza 30, PL-30059 Krak\'ow, Poland

}

\bigskip

$^+${\tt kulakowski@novell.ftj.agh.edu.pl}, $^*${\tt dydejczyk@ftj.agh.edu.pl}

\bigskip

\today

\end{center}

\begin{abstract}
We discuss the simulations of the norm game between players at nodes of a directed random network.
The final boldness of the players can vary with the initial one as the $\Theta$ function. One of the 
conditions of this behaviour is that the player who does not punish automatically becomes a defector. 
The threshold value of the initial boldness can be interpreted as a norm strength. It 
increases with the punishment and decreases with its cost. Surprisingly, it also decreases with the 
number of potential punishers. The numerical results are discussed in the context of the statistical 
data on crimes in Northern Ireland and New Zealand, on divorces in USA, and on the alcohol consumption 
in Poland. 
\end{abstract}





\section{Introduction}

The cultural paradigm of the supremacy of an individual over collective thought is well established in our culture.
On the contrary to this, there is much evidence that individual decisions and beliefs strongly depend
on the social environment. The effects related are the spiral of silence \cite{noe,noelle}, the pluralistic
ignorance \cite{plurig1,plurig2}, the bystander effect \cite{byst}, the group polarization \cite{grpola},
the groupthink \cite{grth}, the bandwagon effect \cite{bandw1,bandw2}, the Abilene paradox \cite{abilene},
stereotyping \cite{lip,sciam} and others. All these effects show that social groups influence or sometimes even
determine their members' behaviour. It is natural that this notion shifts attention of scientists to investigations of 
an interaction between agents rather than the agents themselves, notwithstanding this interaction
perception merely in individual minds. A physicist could be surprised how far the analogy can be driven between the
cognitive-emotional social field \cite{socfield} and the electromagnetic field \cite{socfi}.\\

The subject of social norms seems to belong to these parts of the science on human behaviour, which can be treated
as a systematic science (in terms of Ref. \cite{socfield}). Up to our knowledge, the first attempt of such a treatment
was done by Robert Axelrod \cite{axel1,axel2}. A brief introduction to the problem can be found in the introduction 
to Ref. \cite{fs}. Axelrod was inspired by the idea of the evolutionary principle (EP),
formulated by himself as follows: {\it ...what works well for a player is more likely to be used again, whereas what
turns out poorly is more likely to be discarded.} \cite{axel3}. Axelrod enumerated three independent mechanisms which
lead to the realization of EP: {\it i)} more effective individuals survive more likely, {\it ii)} players learn by
trial and error and maintain strategies which are more effective, {\it iii)} players observe each other and imitate
the winners. Axelrod suggested that in the problem of norms, the third mechanism is most applicable. In the
opinion of the present author the second mechanism is even more important for norms which can persist in the
timescale of the order of human life. The argument is that such norms are characteristic for a social group rather than
for a given person, and they are obeyed sometimes just by tradition. Investigation of this kind of norms with
statistical tools makes sense, as the psychics of an individual agent can be reduced there to a black box. This is
the area where a sociophysicist is allowed to enter.\\

In our previous work \cite{kk} a new scheme of simulation was formulated for the norm game \cite{axel1,axel2}.
Most briefly, the original version of the norm game can be described as follows. $N$ players are positioned at nodes
of a fully connected network. The players are endowed with their boldnesses, i.e. initial probabilities of defecting
(breaking the norm), and with vengeances,
i.e. the probabilities of punishing the defection, if observed at other players. Next, a player is selected and tempted
to break the norm. If he does, he is granted with some prize. Further, if a neigbour of the defector decides to punish,
the defector has to pay a fine. Finally, also the punisher incurs some cost of his action. An application of the
genetic algorithm leads to spreading of most effective strategies.\\

This game, with some simplifications described in Section II, was simulated recently in a
directed Erd\"os-R\'enyi network \cite{kk}. The main result was the history dependence of the final distribution of the
agents' boldness. Namely, a discontinuity was found of the dependence of the final against the initial boldness.
The values of the initial boldness of the players were selected from the range $(0.9 \rho, \rho)$,
where $0<\rho<1$. Below some threshold value $\rho_c$ of $\rho$, the mean value of the final
boldness was found to be close to zero. When $\rho>\rho_c$, the same mean
value was found to be close to one. Calculated values of $\rho_c$ were found to decrease with
the cost parameter $\gamma$. In parallel, a similar bistability was obtained by means of analytical
calculations \cite{kk,jeb}.

The aim of the present work is to investigate this discontinuity. New numerical results to be reported here are as follows: \\
{\it i)} A necessary condition for the appearance of the discontinuity is that once a player abstains from punishing, his boldness is set to one. \\
{\it ii)} The discontinuity remains in the undirected growing networks: both in the scale free and the exponential ones, but it disappears in the 
directed growing networks.\\
{\it iii)} The value of $\rho_c$ increases with the punishment and it decreases with the mean node degree. \\
These results are described in Section III with more details. They are discussed in Section IV in the context of some statistical data, 
which display a relatively quick variations in time. Final remarks in Section V close the text.

\section{The simulation algorithm}

Basically, we use the same algorithm as in Ref. \cite{kk}, generalized for different network topologies.
Players are placed at nodes of a network of a given topology. For the directed Erd\"os-R\'enyi network,
for each node $i$ its degree $k(i)$ is selected from the Poisson distribution, and then the neigbors of $i$
are chosen with uniform probability. The links are directed from the neigbours to the node $k$; this means
that the player at $i$ can be punished by his neighbours, but not the opposite. Obviously, each neigbour has its
own neighbours who can punish him.\\

As the next step, each player $i$ is endowed by the probability that he defects $b(i)$ (boldness) and the probability
that he punishes $v(i)$ (vengeance), with the condition $b(i)+v(i)\le 1$. The boldnesses are selected from the
range $((\rho-1)/100,\rho/100)$, where $\rho$ is an integer number from 1 to 100. Then, $\rho$ is a measure of the
initial boldness in the system. For each $i$, the vengeance is set as $v(i)=(1-b(i))/\mu$, where $\mu\ge 1$.
Then, the initial state is described by two parameters $\rho$ and $\mu$. As a rule, $\mu=1$ or $\mu=2$.\\

The dynamics is described by two parameters $\beta$ and $\gamma$. A node $i$ is selected randomly.
The player at $i$ defects with probability $b(i)$. If he does, his boldness $b(i)$ is set to 1, and his vengeance
$v(i)$ is set to 0. If he does not, $b(i)$ is set to 0 and $v(i)$ is set to 1. This settings can be treated as an 
example of social labelling \cite{ox}. Then, a neighbour $j$ of $i$ is checked
if he punishes or not, with his actual vengeance $v(i)$. If he does not punish, he is treated as a defector.
Then, his boldness $b(j)$ is set to 1 and his vengeance $v(j)$ is set to 0. If $j$ punishes, the boldness
of punished $i$ is reduced by a factor $1-\beta$, i.e.

\begin{equation}
b(i) \to b(i)(1-\beta)
\end{equation}
Simultaneously, the punisher $j$ pays the cost of punishing, what leads to a reduction of his vengeance as

\begin{equation}
v(j) \to v(j)(1-\gamma)
\end{equation}
In this way, we omit the parameters of the payoffs, which are not necessary. If one of the neigbours punishes,
the other neigbours are no more checked if they do. The algorithm just selects another player $i$ to be tempted to defect.\\

The simulation is repeated for the undirected growing networks: exponential and scale-free ones, and for the directed
growing networks. In the last case the direction of nodes was selected either from the new node to its older
neigbours, or the opposite.

\section{Numerical results}

Although the threshold effect is present for the directed Erd\"os-R\'enyi networks \cite{kk}, it disappears in the 
directed Albert-Barab\'asi networks where the direction of bonds is chosen from newly attached nodes to their older neighbours.
The effect disappears also in the latter case when the direction of all bonds is inverted. These results are shown in Fig. 1. 
Then, this combination of the age of nodes and the direction of bonds between them does modify significantly the system behaviour. 
The nature of this modification remains to be clarified.

\begin{figure} 
\vspace{0.3cm} 
{\par\centering \resizebox*{10cm}{8cm}{\rotatebox{-90}{\includegraphics{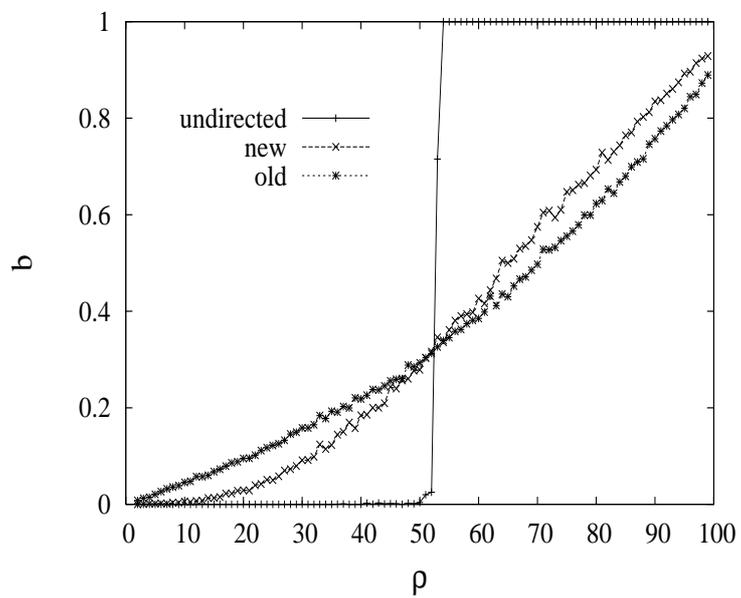}}} \par} 
\vspace{0.3cm} 
\caption{The mean boldness for Albert-Barab\'asi growing networks with $M=3$. For the undirected network the threshold value is observed 
at $\rho_c$ about 53. For two directed networks the direction is either from new to old nodes or the opposite. Then, either new nodes can
punish those to which they are attached (the plot 'new'), or the old nodes can punish those attached to them (the plot 'old'). No threshold
is observed for the directed networks. }
\label{fig-1}
\end{figure}

The simulations were concentrated on the calculation on the threshold boldness $\rho_c$ against the punishment parameter $\beta$
and the cost parameter $\gamma$. Let us repeat that if the initial value of $\rho$, what is a rough measure of the initial
boldness, is smaller than $\rho_c$, then the final boldness is small; the defection is rare. On the contrary, if $\rho > \rho_c$,
most agents defect in the stationary state. In other words, it is more difficult to defect if $\rho_c$ is large. Then, 
$\rho_c$ can be treated as a measure of the norm strength.

\begin{figure} 
\vspace{0.3cm} 
{\par\centering \resizebox*{10cm}{8cm}{\rotatebox{0}{\includegraphics{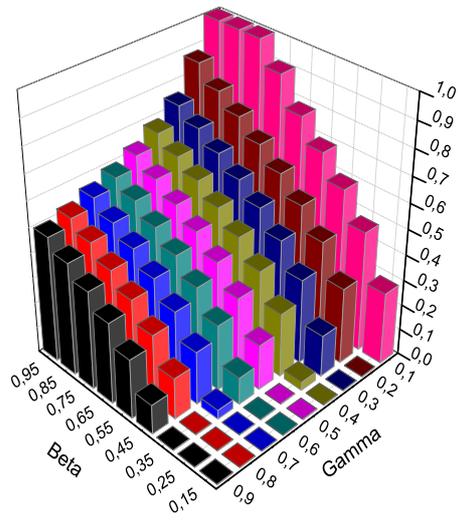}}} \par} 
{\par\centering \resizebox*{10cm}{8cm}{\rotatebox{0}{\includegraphics{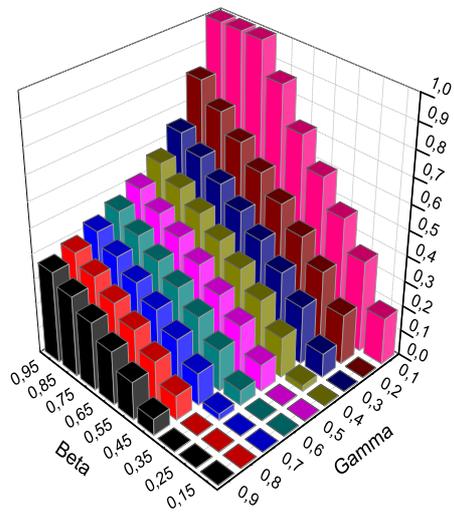}}} \par} 
\vspace{0.3cm} 
\caption{The value of the norm strength $\rho_c$ for the directed Erd\"os-R\'enyi network, for $\mu=1$ and $\mu=2$.}
\label{fig-2}
\end{figure}

The simulations confirm the previous result \cite{kk} that $\rho_c$ increases with the punishment constant $\beta$. Also, they give
a new result, that $\rho_c$ decreases with the punishment cost $\gamma$. Both facts are natural; the norm is stronger if
the punishment is heavy for the defector, and it is weaker if the punishment is costful for the punisher. These results
are shown in Fig. 2 for the directed Erd\"os-R\'enyi network with the mean number of potential punishers $\lambda=5$, and in 
Fig. 3 for the undirected Albert-Bara\'basi networks.

\begin{figure} 
\vspace{0.3cm} 
{\par\centering \resizebox*{10cm}{8cm}{\rotatebox{0}{\includegraphics{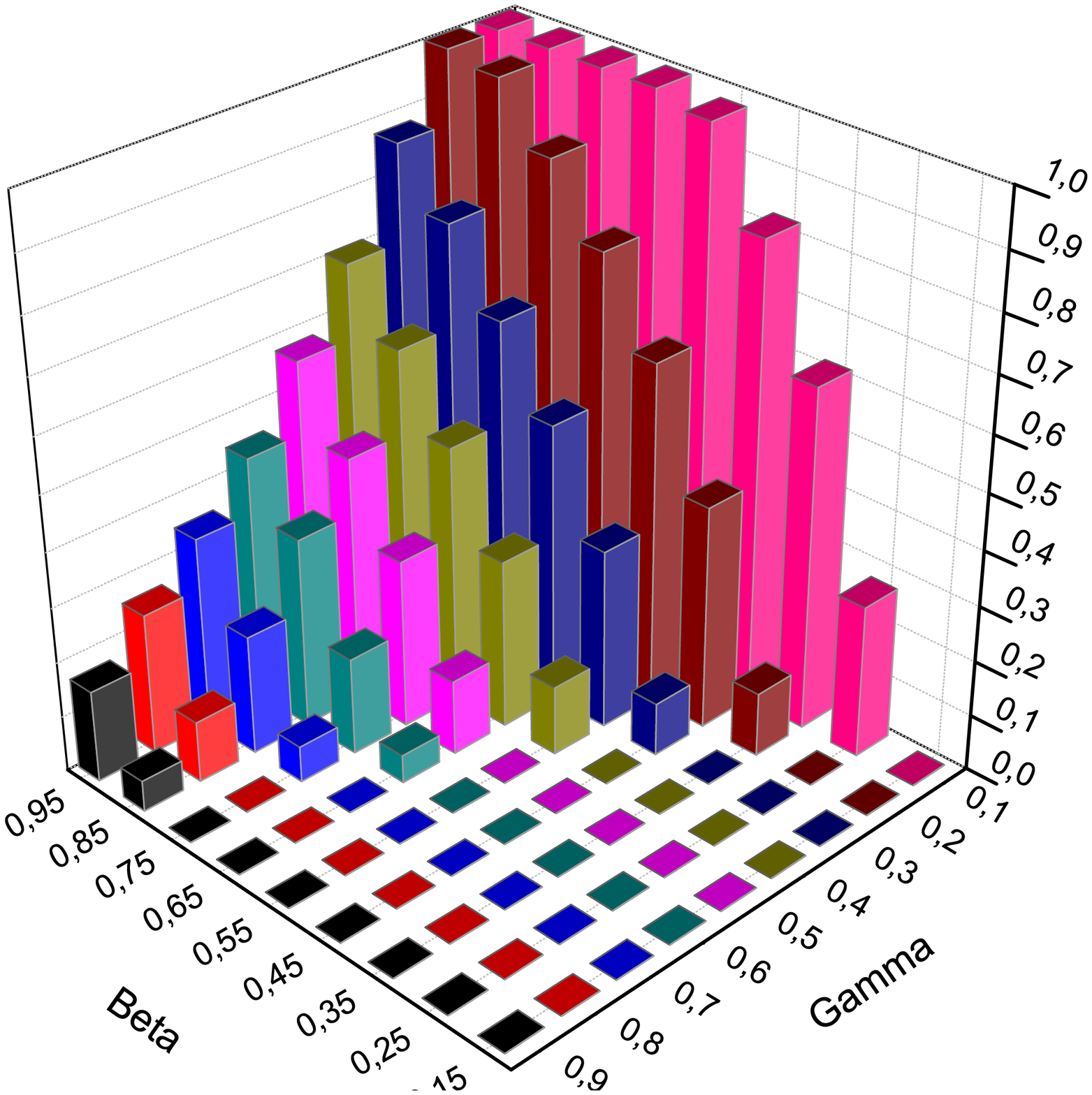}}} \par} 
{\par\centering \resizebox*{10cm}{8cm}{\rotatebox{0}{\includegraphics{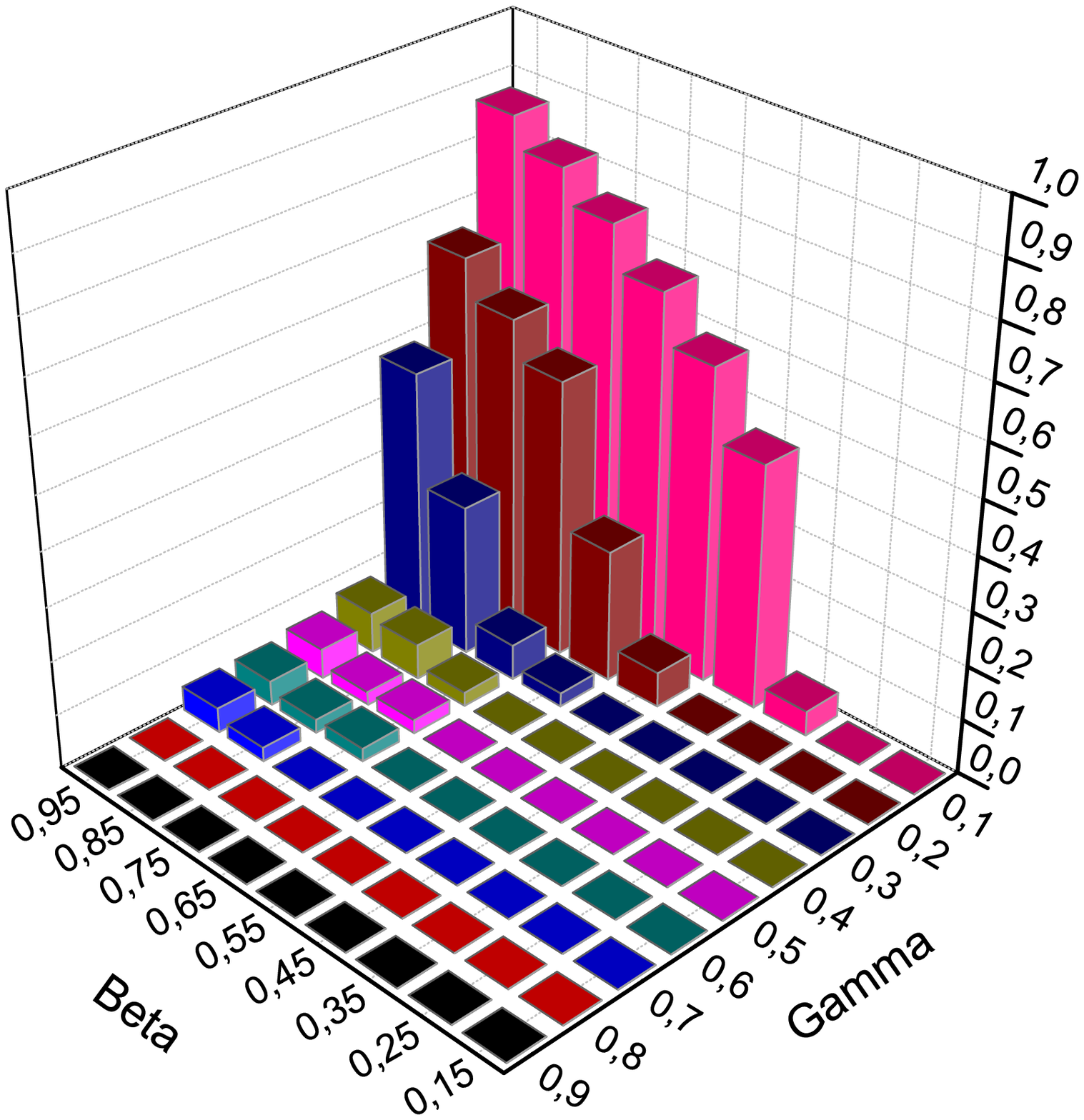}}} \par} 
\vspace{0.3cm} 
\caption{The value of the norm strength $\rho_c$ for the undirected scale-free Bara\'basi-Albert network, for $M=2$ and $M=4$.}
\label{fig-3}
\end{figure}

On the other hand, these results suggest that the norm strength decreases with the potential number of neighbours. 
To verify this, we have calculated $\rho_c$ for the directed Erd\"os-R\'enyi networks for various numbers of in-neigbours $\lambda$ 
as well for the undirected scale-free networks for various numbers of the parameter $M$. The results are shown in Fig. 4 a,b.
In both cases the same effect is observed, that the norm strength $\rho_c$ decreases with the mean number of potential punishers,
which is $\lambda$ for the the directed Erd\"os-R\'enyi networks and which is $2M$ for the undirected scale-free networks.

\begin{figure} 
\vspace{0.3cm} 
{\par\centering \resizebox*{10cm}{8cm}{\rotatebox{-90}{\includegraphics{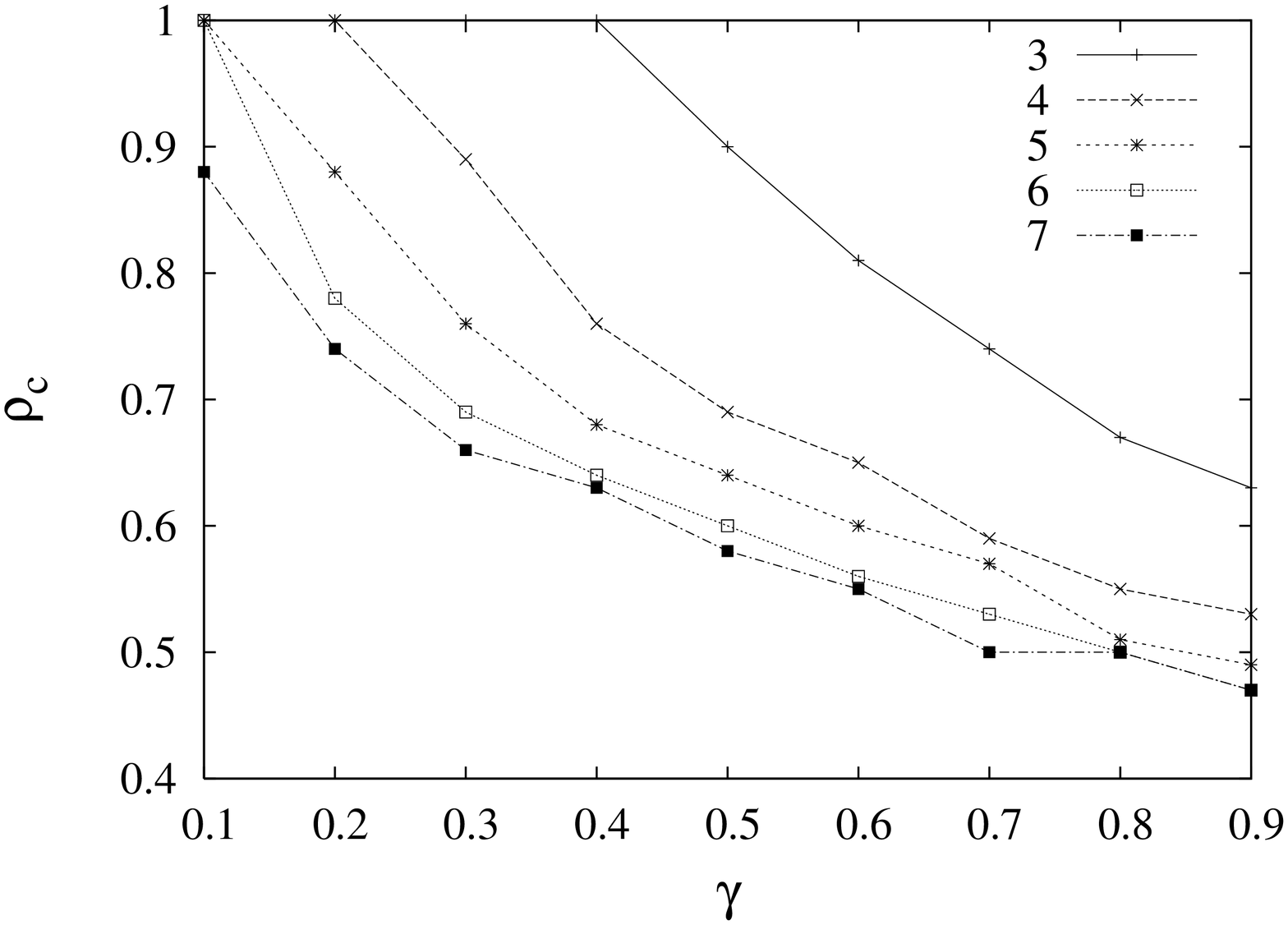}}} \par} 
{\par\centering \resizebox*{10cm}{8cm}{\rotatebox{-90}{\includegraphics{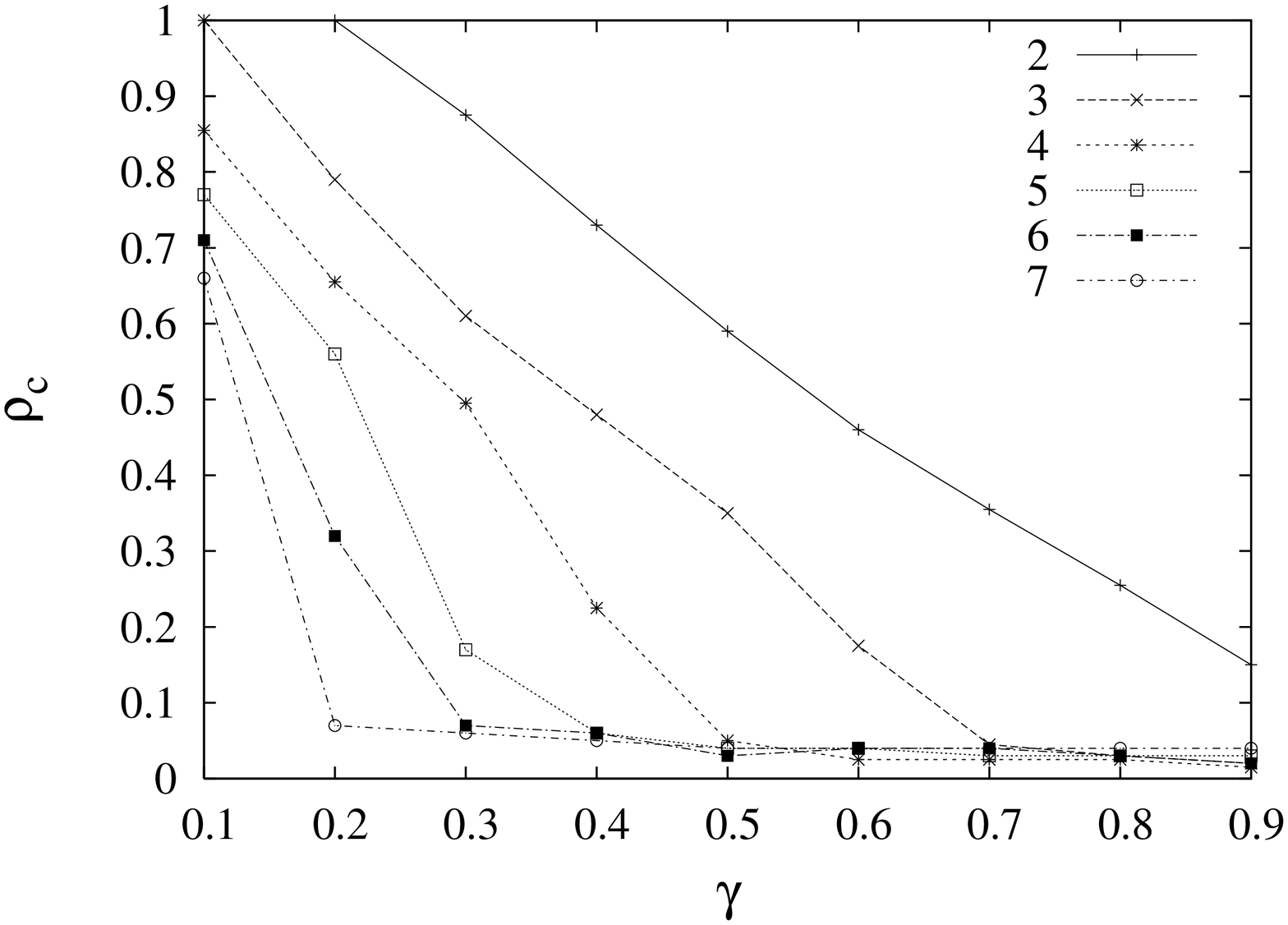}}} \par} 
\vspace{0.3cm} 
\caption{The value of the norm strength $\rho_c$ for the directed Erd\"os-R\'enyi networks for $\lambda=3-7$ and for the undirected scale-free Albert-Barab\'asi network, for $M=2-7$.}
\label{fig-4}
\end{figure}

This results is somewhat surprising. Before the calculation is done, we expected that an increase of the number of punisher should make 
the norm more strong. We have to admit that the process contains some internal feedback from those tempted to their neighbours. Once
the norm is broken, those who can punish have to decide if the punishment cost is not too large. According to our algorithm once they do not 
punish, they become defectors. If the norm was broken at all, this outcome was not possible. In this sense, the larger number of potential 
punishers who do not punish can lead to an avalanche of new defectors. This is a possible explanation of the observed decrease of $\rho_c$
with the number of neighbors.

\section{Some stylized facts}

Some statistical data suggest that indeed some 'equilibrium' social states are possible, where a norm
is obeyed to some definite extent. Once such a state ceases its stability, the system moves to another state,
where the mean boldness is constant again or it fluctuates only slowly. This transformation can be seen
in the statistical data as a large change of a given quantity, accompanied by its relatively slow variation
before and after the change. Such a change can be interpreted as a discontinuous jump between two states.
Here we are going to mention four examples and to comment them shortly.\\

Our first example is the number of recorded criminal offences per year in Northern Ireland. Between 1980 and
1997 it varied between 50 thousands and 70 thousands, but it increased to about 120 thousands in 1997
\cite{ira,nisr}. The plot is shown in Fig. 5. This jump was probably triggered by the fact that since 1997
the Provisional IRA has observed a ceasefire. At this moment, large hidden arsenals became useless for the
patriotic purposes. As a consequence, the social norm 'preserve your gun for the outbreak' has failed.

\begin{figure} 
\vspace{0.3cm} 
{\par\centering \resizebox*{9cm}{7cm}{\rotatebox{0}{\includegraphics{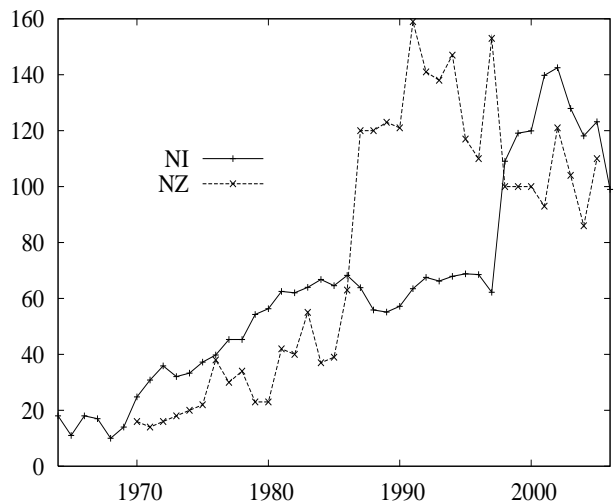}}} \par} 
\vspace{0.3cm} 
\caption{The number of recorded criminal offences in thousands per year in Northern Ireland (+, solid line) \cite{ira,nisr} and the number 
of homicides per year in New Zealand (x, dash line) \cite{nzcrime} against time. Sharp changes are visible, which mark the cease-fire
in Northern Ireland in 1997 and the transition to free economy, started in 1984 in New Zealand.}
\label{fig-5}
\end{figure}

The data on murders in New Zealand between 1950 and now show a similar dynamics \cite{nzcrime}. Before
1986 the number of homicides did not excess 60 per year, but it was not smaller than 120 per year between 1987
and 1994. The plot is shown also in Fig. 5. This change is correlated with the transformation of New Zealand
to free-trade economy, started in 1984. Here the norm is the Commandment 'You shall not murder' which did not
cease its universality in 80-s; perhaps it was the punishment what decreased.\\

Our third example is the number of divorces in USA (Fig. 6). The data \cite{divus} show that this number increased
twice (from about 2.5 to about 5.0 per 1000 population) between 1965 and 1975. This change was accompanied by
a similarly large relative increase of numbers of rapes, murders and robberies within the same time period \cite{uscrime}.
Obviously, this fall of norms was triggered by the Vietnam war, but it is not easy to separate this cause of divorces from the accompanying
sexual revolution. Divorced people often look for other partners, not necessarily unmarried, what accelerates
the process like in a chain reaction \cite{evert}; still the related lifetime is not shorter than a couple of years.\\

\begin{figure} 
\vspace{0.3cm} 
{\par\centering \resizebox*{9cm}{7cm}{\rotatebox{-90}{\includegraphics{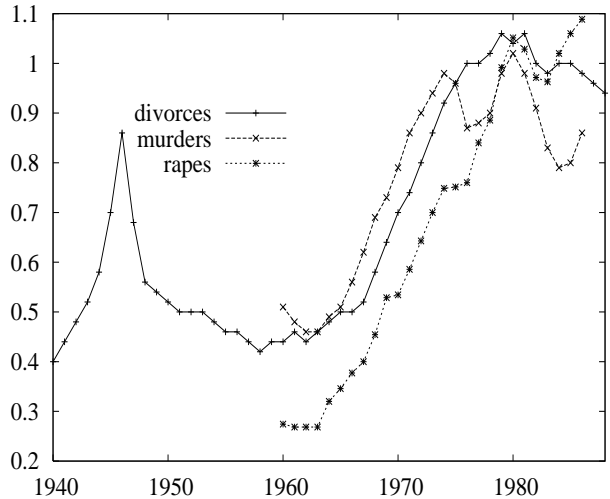}}} \par} 
\vspace{0.3cm} 
\caption{Divorces (+, solid line) \cite{divus}, murders (x, dash line) \cite{uscrime} and rapes (*, doted line) \cite{uscrime} 
in USA against time. For clarity, the numbers are rescaled: divorces to 500, murders to 10 and rapes to 35 per $10^5$ inhabitants 
per year. The peak on the left on the divorces curve is due to WWII. }
\label{fig-6}
\end{figure}

Our last example is the fall of spirits consumption in Poland in 1981 from 8 to 6 litres of pure alcohol per an adult per year
\cite{fao}. At that time, Polish people had drank mainly spirits, while the French preferred wine and the German - beer.
To drink alcohol is in Poland a kind of social norm, to show friendship and empathy, paying no heed to health losses.
This need for a demonstration of common feeling of being unhappy was released by the new hope of a political
transformation, brought by the free trade union "Solidarity" and the message of John Paul II. The political 
transformation in Europe is not visible at the same data for other countries; the consumption of wine in France and of beer 
in Germany decreased rather smoothly \cite{fao}, as shown in Fig. 7.\\

As we see, all these changes were triggered by some events: the ceasefire in Northern Ireland, the economic
transformation in New Zealand, Vietnam war in USA or political reforms in Poland. To interpret these changes
in terms of our model picture, we admit that in each case the event as those listed above changes the punishment or/and
it cost. This influences the temptation process, which is more or less steady in the society; new and new people are 
faced with the question: to divorce or not, to drink or not, to break law or not. Their decisions are influenced by individual 
attitudes of their neighbours, which are also determined in the process. As an output we see a social change in macroscopic data.\\

\begin{figure} 
\vspace{0.3cm} 
{\par\centering \resizebox*{9cm}{7cm}{\rotatebox{-90}{\includegraphics{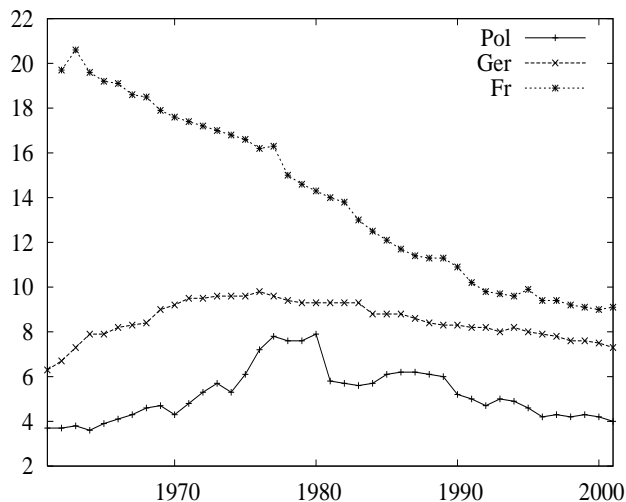}}} \par} 
\vspace{0.3cm} 
\caption{Consumption of spirits in Poland (+, solid line), beer in Germany (x, dash line) and wine in France (*, dotted line), 
in litres of pure alcohol per an adult (+15) per year, against time. The consumption of beer and wine in Poland, wine and 
spirits in Germany and spirits and beer in France did not excess 4 litres. All data from Ref. \cite{fao}.}
\label{fig-7}
\end{figure}

We note that according to our numerical results, the abrupt fall of norm preserving is a consequence of the rule
'those who abstain from punishing, defect'. In other words, the society is divided into defectors and punishers.
A question arises, if this rule makes sense in our four examples. To find an answer, we have to identify the punishment.
In the first and second example to punish is to report the case to the police. Then, the interpretation is
clear: who did not report to the police is guilty. In the third example to punish could be to condemn divorces in public.
However, in principle one could be faithful in marriage and tolerant towards manifestations of sexual promiscuity.
We face the same problem also in the fourth example, where many people prefer the strategy 'don't drink but let drink'.
Then, our numerical result seems to apply to two examples out of four.\\

This difficulty is solved, if the norm itself is redefined as 'the duty to punish'. According to this new point of view,
to obey a norm means to declare in public that this norm should be obeyed. In particular in our third example what is
important is to declare in public that the institution of marriage is holy and inviolable. Similarly in the fourth example
the norm is to insist that to be accepted in the group, everybody must drink. In this way, the metagame (in terms of
Axelrod) is the game. This transformation can be interpreted theoretically as follows: the preservation of a given norm 
should be evaluated by counting not those people who preserve it, but rather those who punish for its defection. This 
formulation raises the question about the status of norms which vanish in a continuous way. This in turn calls for an 
experimental criterion, which change is continuous and which is sharp. These questions are beyond the scope of this paper.\\

\section{Final remarks}

Even if numerically complete, the study is preliminary. It reveals the connection between consequences of decisions 
of individual players and the state of the society at a macro-level. In the above calculations, these consequences are 
chosen as to label the players: who once defected will never punish and who once punished will never defect.
In this way, the society is divided into two opposite groups. These rules of labelization should be justified and relaxed 
if necessary in accordance with the social reality, and these modifications in each case do depend on the specific norm under 
consideration. \\

To conclude, the sharp dependence of the final boldness on its initial distribution follows from the labelling consequences
of initial decisions of the players. This sharpness does not depend much on the network topology. However, it can
be removed if the direction of the links is determined by the age of nodes in growing networks.\\

\bigskip

{\bf Acknowledgements.} The research is partially supported within the FP7 project SOCIONICAL, No. 231288.

\end{document}